%% file: main.tex
\documentclass[10pt,twocolumn,letterpaper]{article}

\usepackage{cvpr}              
\usepackage{multirow, colortbl, bm, arydshln}

\definecolor{cvprblue}{rgb}{0.21,0.49,0.74}
\usepackage[pagebackref,breaklinks,colorlinks,allcolors=cvprblue]{hyperref}
\usepackage[accsupp]{axessibility}

\hypersetup{
    colorlinks=true,    
    linkcolor=red,     
    citecolor=green,     
    urlcolor=magenta        
}

\def\paperID{9585} 
\def\confName{ICCV}
\def\confYear{2025}

\title{Physical Degradation Model-Guided Interferometric Hyperspectral Reconstruction with Unfolding Transformer}

\author{Yuansheng Li\textsuperscript{1} \quad Yunhao Zou\textsuperscript{1} \quad Linwei Chen\textsuperscript{1} \quad Ying Fu\textsuperscript{1}\thanks{Corresponding author: fuying@bit.edu.cn} \\
\textsuperscript{1}Beijing Institute of Technology\\
{\tt\small 2578548450@qq.com \quad \{wangyuran,liangyingping,fuying\}@bit.edu.cn}
}

\begin{document}
\maketitle
\input{sec/_0_abstract}    
\input{sec/_1_introduction}

\input{sec/_2_related_works}
\input{sec/_3_methodology}
\input{sec/_4_experiments}
\input{sec/_5_conclusion}

\section*{Acknowledgements}
This work was supported by the National Key R\&D Program of China (2022YFC3300705), the National Natural Science Foundation of China (62331006, 62171038, and 62088101), and the Fundamental Research Funds for the Central Universities.

{
    \small
    \bibliographystyle{ieeenat_fullname}
    \bibliography{main}
}


\end{document}

%% file: sec/_0_abstract.tex
\begin{abstract}
    \indent
    Interferometric Hyperspectral Imaging (IHI) is a critical technique for large-scale remote sensing tasks due to its advantages in flux and spectral resolution. However, IHI is susceptible to complex errors arising from imaging steps, and its quality is limited by existing signal processing-based reconstruction algorithms. Two key challenges hinder performance enhancement: 1) the lack of training datasets. 2) the difficulty in eliminating IHI-specific degradation components through learning-based methods.  To address these challenges, we propose a novel IHI reconstruction pipeline. First, based on imaging physics and radiometric calibration data, we establish a simplified yet accurate IHI degradation model and a parameter estimation method. This model enables the synthesis of realistic IHI training datasets from hyperspectral images (HSIs), bridging the gap between IHI reconstruction and deep learning. Second, we design the \textbf{I}nterferometric \textbf{H}yperspectral \textbf{R}econstruction \textbf{U}nfolding \textbf{T}ransformer (IHRUT), which achieves effective spectral correction and detail restoration through a stripe-pattern enhancement mechanism and a spatial-spectral transformer architecture. Experimental results demonstrate the superior performance and generalization capability of our method. The code and are available at \href{https://github.com/bit1120203554/IHRUT}{https://github.com/bit1120203554/IHRUT}.
\end{abstract}

%% file: sec/_1_introduction.tex
\section{Introduction}
\label{sec:intro}

In the field of remote sensing, hyperspectral images (HSIs) serve as crucial repositories of spectral information related to materials and structures. Among various imaging techniques, Interferometric Hyperspectral Imaging (IHI) captures interferograms using instruments and reconstructs HSIs through algorithms based on the spectral Fourier Transform (FT) relationship between interferograms and HSIs \cite{FTS,FTS2}. Benefiting from its strengths in flux and spectral resolution \cite{fellgett,jacquinot}, IHI is extensively utilized in large-scale tasks aboard spacecraft \cite{application1}.

\begin{figure}[t]
    \centering
    \includegraphics[width=\linewidth]{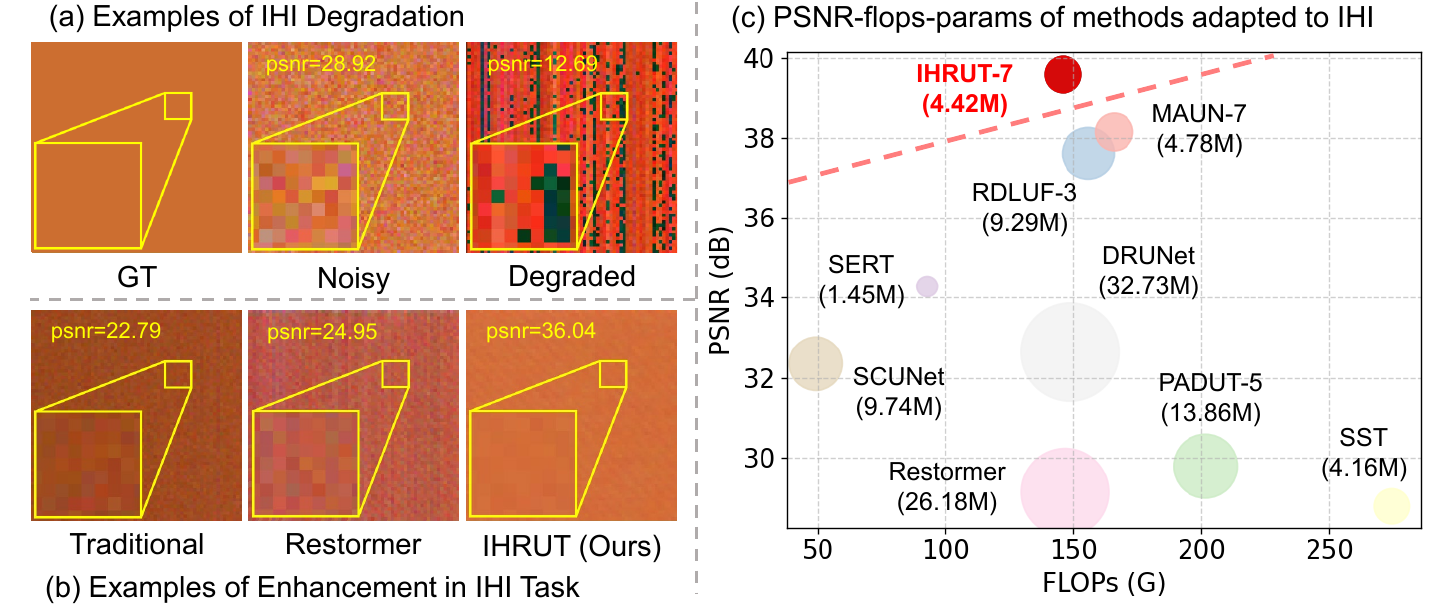} 
    \caption{Examples of degradation and reconstruction of IHI. (a) Spectral RGB visualization of IHI calibration data across three states: ideal, corrected yet noisy, and degraded, illustrating degradation components such as color shifts and stripe patterns in IHI. (b) Comparison of the same data processed by traditional methods, Restormer, and IHRUT trained on synthetic dataset. IHRUT outperforms the traditional method and the direct learning approach of Restormer. (c) Comparison of PSNR and complexity of learning-based methods highlights IHRUT's superior performance in IHI reconstruction.}
    \label{fig:1}
    \phantomsubcaption\label{fig:1a}
    \phantomsubcaption\label{fig:1b}
    \phantomsubcaption\label{fig:1c}
\end{figure}

Despite its advantages, IHI faces significant degradation issues arising from frequent overexposure and low-light conditions in satellite-borne scenarios, as well as complex imaging steps including scanning, interference, and sensing, as depicted in Fig. \ref{fig:2b}. Imaging degradation leads to complex error distributions \cite{IHIdegradation} and significant signal loss (in Fig. \ref{fig:1a}), posing crucial challenges for reconstruction. However, existing IHI reconstruction methods are constrained in performance (as seen in Fig. \ref{fig:1b}) and efficiency by traditional frameworks and time-consuming correction steps. Although deep learning methods, including end-to-end networks and model-assisted networks such as Plug and Play \cite{PnP} and deep unfolding \cite{DAUHST,PADUT,RDLUF}, show effectiveness in HSI reconstruction, these methods remain underutilized in IHI tasks. To break the limitations of conventional reconstruction frameworks and enhance the imaging quality of IHI, two primary challenges need to be addressed sequentially. 

The first challenge lies in the lack of datasets for IHI, which hinders the application of learning-based methods in IHI. The high cost of data acquisition and significant hardware dependency limit the scale and generalizability of real IHI datasets for training. In the absence of real data, generating synthetic datasets from specific degradation models \cite{Hetero,eld1,zhangtao1} is a common approach. Nevertheless, existing degradation models in IHI radiometric calibration \cite{calibration1,calibration2} are primarily based on complex optical processes and are not suitable for data synthesis. Effective simulation methods for IHI are still lacking.

The second challenge lies in addressing special components of IHI degradation during the adaptation of learning-based methods. Although deep reconstruction networks possess powerful representation capabilities, they face difficulties in addressing specific characteristics of IHI degradation through direct learning without priors. For instance, in Fig. \ref{fig:1c}, stripe patterns caused by the scanning mechanism of IHI are difficult to eliminate through global spectral processing and denoising in Restormer \cite{Restormer}, shown as the vertical stripe noises in the result. To better address degradation, priors and optimizations targeting these characteristics are necessary for reconstruction networks.

In this paper, we propose a novel method for IHI reconstruction. To address the lack of datasets for learning, we propose a novel IHI data synthesis pipeline. Specifically, we establish a practical and accurate degradation model adaptable to data synthesis by simplifying the imaging physics through matrix operations. Based on the calibration and solution of the proposed model, we provide a method for generating realistic IHI datasets from HSIs. Furthermore, based on the generated dataset, we propose an \textbf{I}nterferometric \textbf{H}yperspectral \textbf{R}econstruction \textbf{U}nfolding \textbf{T}ransformer (IHRUT) optimized for addressing the IHI degradation components. Instead of direct learning without priors, we introduce a deep unfolding architecture and integrate the degradation model to guide the reconstruction process. Focusing on the specific characteristics of degradation, we introduce a stripe-pattern enhancement mechanism for an adaptive spatial-spectral transformer network. IHRUT exhibits low complexity and outstanding enhancement performance in IHI reconstruction, as illustrated in Fig. \ref{fig:1c}. Our contributions are summarized as follows:

\begin{itemize}
    \item We establish an accurate and simplified IHI degradation model based on imaging physics and experiments with radiometric calibration data.
    \item We propose a calibration-based model parameter estimation and data simulation pipeline to extract key features of degradation and generate realistic IHI datasets.
    \item We propose the IHRUT, an unfolding transformer network for IHI reconstruction tasks, which is verified to be effective and lightweight in reconstruction experiments.
\end{itemize}

%% file: sec/_2_related_works.tex
\section{Related Works}
\label{sec:related}
  
In this section, we review the researches most relevant to our work, including IHI, along with methods for dataset synthesis and HSI reconstruction.

\subsection{Interferometric Hyperspectral Imaging}

IHI stems from the FTS theories {\cite{FTS,FTS2,fellgett}} and is composed of two parts as imaging and reconstruction. 
For imaging, existing IHI instruments are categorized into temporal, spatial, and spatio-temporal modulated types, each with distinct imaging processes, such as TS-SHIS {\cite{TS_SHIS}} and LASIS {\cite{LASIS}}. For reconstruction, existing algorithms primarily consist of the following steps: correction for bad pixel {\cite{badpixel1,IHIcorrection1}}, trend terms {\cite{Trend1}}, and phase {\cite{Mertz,Forman}}, followed by apodization {\cite{Apodization1,Apodization2}}, FTS transformation and post-processing. Despite these model-driven methods, Chen et al. {\cite{FCUN}} recently attempted learning-based FCUN for reconstruction. However, FCUN is limited in reconstruction of pixel without considering spatial domain characteristics of HSIs. The lack of data restricts the development of learning-based IHI reconstruction with superior performance.

\subsection{Degradation Modeling and Dataset Synthesis}

Learning-based restoration {\cite{dianzi1,zhang2025unaligned,dianzi4}} approaches typically draw their potential from data. However, the acquisition of large-scale real datasets is often challenging, and one solution is using degradation models as priors to generate synthetic datasets {\cite{Li_2025_CVPR}}. Common models includes Gaussian white noise and homoscedastic Gaussian noise {\cite{Gaussian1,Gaussian2,Hetero}}, among others. Wei {\cite{eld1,eld2}} introduced a camera calibration-based, precise noise model that allows simulated data to yield low-light noise recovery akin to real-data training outcomes. Zhang {\cite{zhangtao1,zhangtao2,dianzi2}} extended this to HSI denoising, presenting a realistic dataset and an HSI noise model. These methods provide inspirations for addressing the challenges in existing IHI calibration and generating training data.

\subsection{Hyperspectral Image Reconstruction}

Existing HSI reconstruction focus on tasks such as RGB or infrared-to-HSI conversion {\cite{gao2025grayscale,RGB2HSI}}, and compressed sensing imaging (SCI) \cite{CASSI,dianzi3}. Deep learning approaches outperforms traditional model-driven methods {\cite{TwIST,GAPTV}} in terms of performance and generalization. These methods are categorized into two types: end-to-end (E2E) and model-aided networks. E2E methods directly learn the reconstruction mapping , with examples including CNN-based TSA-Net {\cite{TSANet}} and HDNet {\cite{hdnet}}, RNN-based BIRNAT {\cite{BIRNAT}}, and transformer-based MST, MST++ , and CST {\cite{mst,mst++,cst}}. Model-aided networks, exemplified by Deep Plug-and-Play (PnP) {\cite{PnP}} and Deep Unfolding (DU), integrate iterative optimization with deep learning. Among these, DU adopts the same training strategy as E2E methods, thereby enhancing the network's adaptability to iterations compared to PnP, and achieving superior restoration performance. Examples include GAP-Net {\cite{GAPNet}}, transformer-based DAUHST {\cite{DAUHST}}, Proximal Gradient Descent (PGD)-based RDLUF {\cite{RDLUF}}, and Memory-Augmented MAUN {\cite{MAUN}}. However, for IHI tasks, these algorithms have yet to be applied and optimized.

\begin{figure*}
    \centering
    \includegraphics[width=0.95\textwidth]{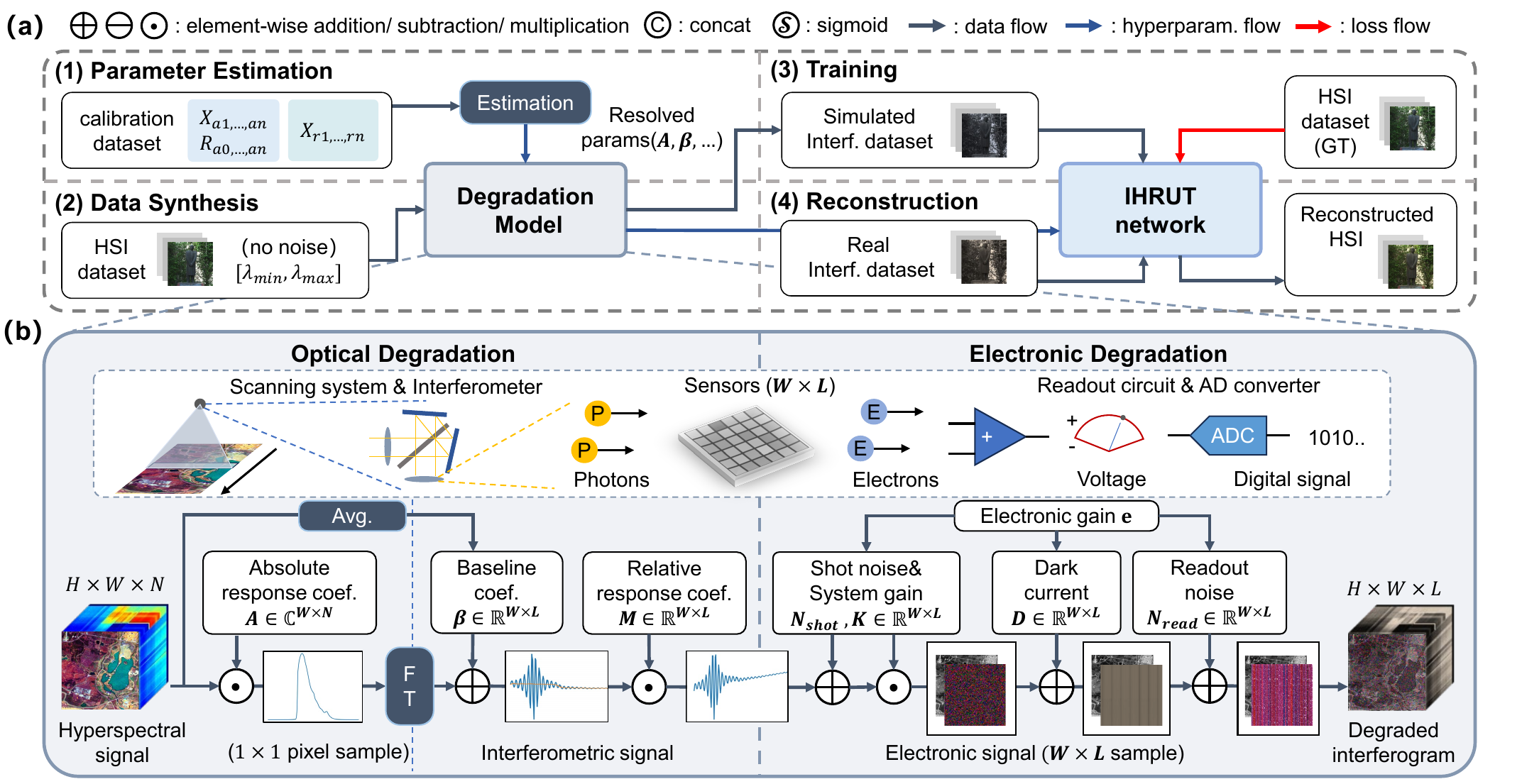}
     \caption{(a) Our overall research framework. We estimate the parameters of the degradation model and incorporate HSI to generate synthetic dataset. Then, we train IHRUT on the synthetic dataset to endow it with the capability of real IHI reconstruction. (b) The degradation model is divided into 2 stages: electronic and optical degradation.
     }
     \label{fig:2}
     \phantomsubcaption\label{fig:2a}
     \phantomsubcaption\label{fig:2b}
     \phantomsubcaption\label{fig:2c}
  \end{figure*}


%% file: sec/_3_methodology.tex
\section{Method}
\label{sec:method}

IHI is essentially a hardware-encoded, software-decoded process. Given a scene \( \mathbf{C} \), the observed image \( \mathbf{I}_d \) and reconstructed HSI \( \mathbf{C}' \) are obtained by:
\begin{equation}
  \begin{aligned}
    \mathbf{I}_d = G_{\Omega}(\mathbf{C}), \; \mathbf{C}' = G'(\mathbf{I}_d),
  \end{aligned}  
  \label{eq:1}
\end{equation}
where \( G_{\Omega} \) is the imaging process and \( G' \) is the IHI reconstruction algorithm. Our research focuses on optimizing \( G' \), with the overall architecture shown in Fig.~\ref{fig:2a}. We model \( G \) into a simplified degradation model with calibration methods, and generate simulated data from the model. To overcome the limitations of \( G' \), we design and train IHRUT tailored to the degradation characteristics, achieving performance optimization.

\subsection{Degradation Modeling}
\label{sec:3.1}

Our degradation model is based on LASIS~\cite{LASIS}, a scanning-based interferometric imaging spectrometer. The size of the sensor array in LASIS is \( W \times L \), where each unit is responsible for scanning a column (\( H \times 1 \times 1 \)) of the interferogram with a specific optical path difference (OPD). The errors are highly correlated within each signal column, but vary significantly between different columns depending on the response capabilities of sensors. We refer to this characteristic as the \textbf{stripe pattern}, which serves as a basic assumption of our degradation model. The degradation model is divided into two stages: \textbf{optical} and \textbf{electronic}.

\noindent \textbf{Optical Degradation.}
Optical degradation corresponds to various error terms during the propagation of light through the optical path of interferometer into the sensors. 

The input light signal can be considered as a hyperspectral signal in a certain wavelength range \( [\lambda_{\text{min}}, \lambda_{\text{max}}] \). Assuming expression of the signal in the form of wavenumber $\nu$ ($\nu = 1/ \lambda \in [1/\lambda_{\text{max}}, 1/\lambda_{\text{min}}] $) is \( \mathbf{B}_0(\nu) \in \mathbb{R}^{H \times W \times N} \). According to theories of IHI~\cite{FTS}, \( \mathbf{B}_0 \) is Fourier transform-related to the interferogram. However, due to the imaging system's varying sensitivity to light in different frequencies, the actual captured spectral signal is the result of the original signal multiplied by a coefficient vector \( \mathbf{A} \) in the spectral domain. Additionally, pose deviations in the interferometer lead to asymmetry in the interferogram, introducing a complex component as phase error into \( \mathbf{A} \). We define \( \mathbf{A} \in \mathbb{C}^{W \times N} \) as the \textbf{absolute response coefficient}, and the interferogram signal can be represented as:
\begin{equation}
  \mathbf{I}_1(l) = \mathcal{F}\{\mathbf{A}(\nu) \odot \mathbf{B}_0 \},
  \label{eq:2}
\end{equation}
where \( \odot \) is element-wise multiplication, and \( l \) denotes OPD.

In the process of Eq. {\ref{eq:2}}, $\mathbf{I}_1$ is still approximately zero-centered. Nevertheless, the actual interferogram is superimposed with background light on each pixel, rendering all pixel values non-negative. The background light can be regarded as a partial component of the incident light, and its value is proportional to the average spectral brightness of the pixels in $\mathbf{B}_0$. Finally, when the interferogram signal containing background light propagates to the sensor, the actual interferogram signal on the image plane is non-uniform due to instrument errors, differences in sensor units, and the frame transfer effect of the CCD. This non-uniformity disrupts the smoothness of the interferograms in the $W$ and $L$-directions, as shown in Fig. {\ref{fig:2b}}. Let $\bm{\beta} (\nu) \in \mathbb{R} ^ {W \times L}$ denote the \textbf{background coefficient}, and $\mathbf{M}\in \mathbb{R} ^ {W \times L}$ denote the non-uniformity as \textbf{relative response coefficient}, the overall model for optical degradation is:
\begin{equation}
  \begin{aligned}
    \mathbf{I}_O &= \mathbf{M} \odot (\mathbf{I}_1 + \bm{\beta} \odot \mu_N(\mathbf{B}_0))\\
    &= \mathbf{M} \odot (\mathcal{F}\{\mathbf{A} \odot \mathbf{B}_0 \} + \bm{\beta} \odot \mu_N(\mathbf{B}_0)),
  \end{aligned} 
  \label{eq:3}
\end{equation}
where \( \mu(\cdot) \) denotes the mean value and \( \sigma(\cdot) \) denotes the standard deviation. 

\noindent \textbf{Electronic Degradation.} 
Electronic Degradation refers to a series of random noises introduced during the signal conversion process after the light enters the sensors.

After entering the sensor, the light signal undergoes the photoelectric effect. Due to the quantum property of light, the number of photoelectrons collected by the sensor within the exposure time is inevitably uncertain and follows a Poisson distribution, introducing \textbf{shot noise}, as:
\begin{equation}
  \begin{aligned}
    &\mathbf{I}_2 = \mathbf{K} \odot (\mathbf{I}_O + \mathbf{N}_{\text{shot}}), &\\
    &(\mathbf{I}_O + \mathbf{N}_{\text{shot}}) \sim \mathcal{P}(\mathbf{I}_O),&
  \end{aligned}
  \label{eq:4}
\end{equation}
where \( \mathcal{P}(\cdot) \) denotes the Poisson distribution, \( \mathbf{K} \in \mathbb{R}^{1 \times W \times L} \) is the \textbf{system gain} related to the sensitivity (ISO) of the sensor array. The amplitude of shot noise is dependent on the signal \( \mathbf{I}_O \).

Subsequently, the photoelectrons are converted into voltage signals via the readout circuit of the CCD sensor and then digitized through analog-to-digital conversion. In this process, a series of signal-independent degradations are introduced in the form of non-zero mean noise. In the noise, \textbf{dark current} \( \mathbf{D} \in \mathbb{R}^{W \times L} \) is the main non-zero mean component, while zero-mean \textbf{readout noise} \( \mathbf{N}_{\text{read}} \) constitutes the remaining part. The amplitude of \( \mathbf{N}_{\text{read}} \) formed by different sensor units varies significantly due to the stripe pattern characteristic, but their normalized results are approximately uniform Gaussian noise. Let \( \bm{\sigma}_{\text{read}} \in \mathbb{R}^{W \times L} \) denote the amplitude of readout noise on the sensor array, then the full model of electronic degradation can be expressed as:
\begin{equation}
  \begin{aligned}
    &\mathbf{I}_d = \mathbf{K} \odot (\mathbf{I}_O + \mathbf{N}_{\text{shot}}) + \mathbf{D} + \mathbf{N}_{\text{read}}, &\\
    &\mathbf{N}_{\text{read}} \sim \mathcal{N}(0, \bm{\sigma}_{\text{read}}),&
  \end{aligned}
  \label{eq:5}
\end{equation}
where \( \mathbf{I}_d \) is the degraded interferogram according to Eq.~\ref{eq:1}, and \( \mathcal{N}(\mu, \sigma) \) denotes the Gaussian distribution.

\subsection{Calibration and Simulation}
\label{sec:3.2}

\begin{figure*}
  \centering
  \includegraphics[width=0.95\textwidth]{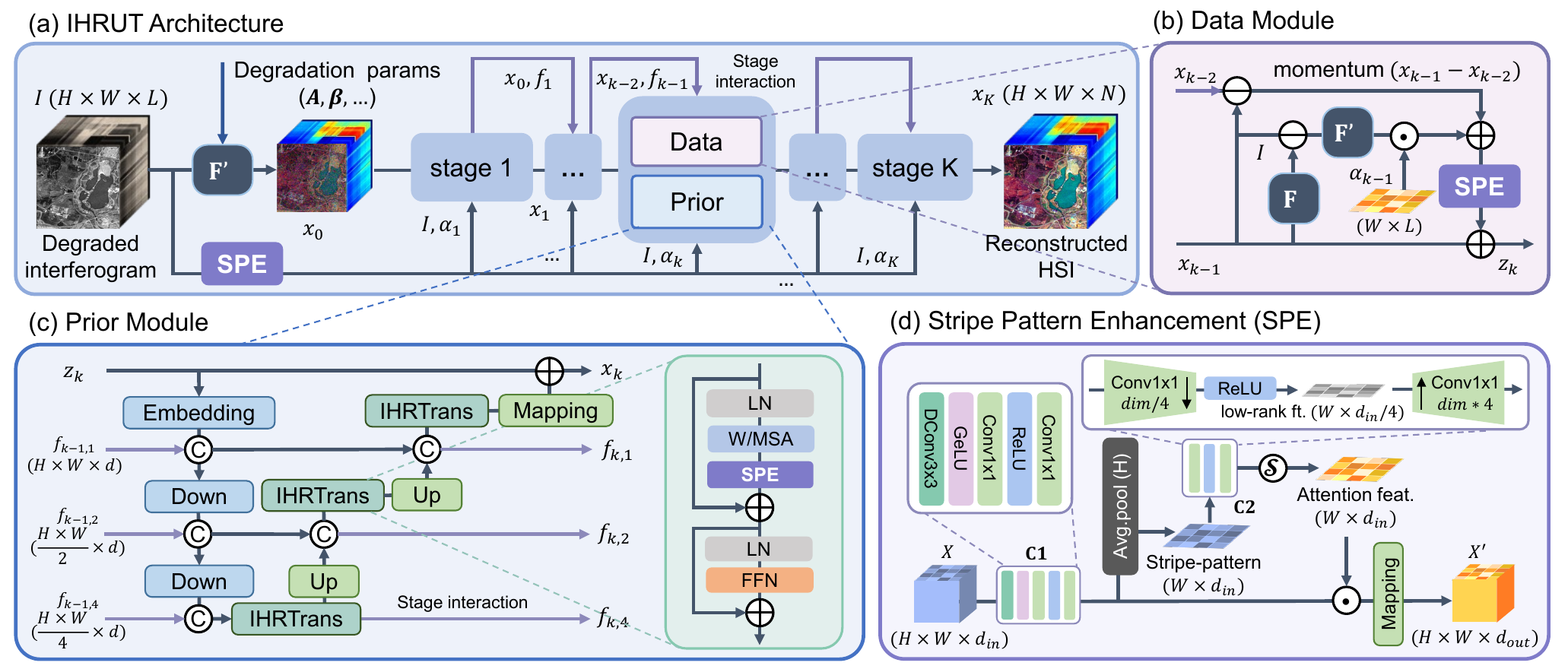}
   \caption{The Architecture of IHRUT network. (a) The overall unfolding framework (b) The stripe-adaptive momentum Data module (c) The lightweight Prior module with stage interaction (d) The Stripe Pattern Enhancement (SPE) module, which is the core of IHRUT.}
   \label{fig:3}
   \phantomsubcaption\label{fig:3a}
   \phantomsubcaption\label{fig:3b}
   \phantomsubcaption\label{fig:3c}
   \phantomsubcaption\label{fig:3d}
\end{figure*}

Parameter estimation involves separation and extraction of degradation components from the radiometric calibration data captured by the LASIS instrument. The calibration data consists of two parts: the absolute calibration data \( \mathbf{I}_A \), which includes uniform light interferograms \( \mathbf{I}_{Ai} \) (\( i = 1, \dots, n \)) with varying brightness levels, along with their corresponding spectra \( \mathbf{B}_{Ai} \) (\( i = 1, \dots, n \)) (expressed in wavenumbers). The relative calibration data \( \mathbf{I}_R \) is captured with an interferometer-removed instrument, including a dark measurement \( \mathbf{I}_{R0} \) and a series of uniform light measurements \( \mathbf{I}_{Ri} \) (\( i = 1, \dots, n \)) without interference patterns.

The parameters for electronic degradation can be derived from \( \mathbf{I}_R \). The dark current \( \mathbf{D} \) is obtained by calculating the mean value of \( \mathbf{I}_{R0} \) along the \( H \)-direction, while the standard deviation of the remaining noise yields \( \bm{\sigma}_{\text{read}} \). Subsequently, compute the mean and variance for each \( \mathbf{I}_{Ri} \) along the \( H \)-direction as \( \mu_H(\mathbf{I}_{Ri}) \) and \( \sigma^2_H(\mathbf{I}_{Ri}) \). By eliminating the signal-independent component (\( \mathbf{D} \) and \( \bm{\sigma}^2_{\text{read}} \)), the mean and variance of the signal-dependent component are estimated as \( \mu_{\text{shot}} \) and \( \bm{\sigma}_{\text{shot}} \). Since the Poisson distribution exhibits the property that the variance equals the mean (\( \sigma^2(\mathbf{N}_{\text{shot}}) = \mathbf{I}_O \)), \( \mathbf{K} \) can be derived from the ratio of \( \bm{\sigma}_{\text{shot}} \approx \mathbf{K}^2 \odot \sigma^2(\mathbf{N}_{\text{shot}}) \) to \( \bm{\mu}_{\text{shot}} \approx \mathbf{K} \odot \mathbf{I}_O \), as:
\begin{equation}
  \begin{aligned}
    &\mathbf{D} = \mu_H(\mathbf{I}_{R0}), \bm{\sigma}_{\text{read}} = \sigma_H(\mathbf{I}_{R0}),&\\
    &\mathbf{K}_i = \frac{\bm{\sigma}_{\text{shot}}}{\bm{\mu}_{\text{shot}}} = \frac{\sigma^2_H(\mathbf{I}_{Ri}) - \bm{\sigma}_{\text{read}}^2}{\mu_H(\mathbf{I}_{Ri}) - \mathbf{D}}.&
  \end{aligned}
  \label{eq:6}
\end{equation}

The parameters for optical degradation are estimated after eliminating electronic degradation components from \( \mathbf{I}_R \) and \( \mathbf{I}_A \). For modulation factor \( \mathbf{M} \), we divide it into two components as \( \mathbf{M} = \mathbf{M}_R \odot \mathbf{M}_A \). \( \mathbf{M}_R \) is caused by sensor non-uniformity and is extracted from \( \mu_H(\mathbf{I}_R) \), while \( \mathbf{M}_A \) arises from distortions in the scanning process and can be obtained through frequency domain analysis of \( \mu_H(\mathbf{I}_A) \). Subsequently, we obtain \( \mathbf{I}_{Ai}' \) by removing \( \mathbf{M} \) and \( \mathbf{D} \) from \( \mu_H(\mathbf{I}_{Ai}) \). The parameter \( \mathbf{A} \) can be derived by comparing the wavenumber representations of \( \mathbf{B}_{Ai} \) and \( \mathbf{I}_{Ai}' \), while \( \bm{\beta} \) can be extracted from the portions of \( \mathbf{I}_{Ai}' \) outside the main signal bands in \( \mathbf{B}_{Ai} \), as:
\begin{equation}
  \begin{aligned}
    &\mathbf{I}_{Ai}' = (\mu_H(\mathbf{I}_{Ai}) - \mathbf{D}) / \mathbf{M}, \; \mathbf{A}_i = \mathbf{B}_{Ai} / \mathcal{F}\{\mathbf{I}_{Ai}'\},&\\
    &\bm{\beta}_i = (\mathbf{I}_{Ai}' - \mathcal{F}\{\mathbf{B}_{Ai}\}) / \mu_N(\mathbf{B}_{Ai}).&
  \end{aligned}
  \label{eq:7}
\end{equation}

During the actual process of parameter selection, relatively accurate values of \( \mathbf{K} \), \( \mathbf{M} \), \( \mathbf{A} \), and \( \bm{\beta} \) can be obtained by averaging the estimations from multiple \( i \). Additionally, we observed that the parameters for electronic degradation exhibit randomness depending on the state of the sensors, while their logarithms demonstrate a linear correlation~\cite{eld2}. To enhance the robustness of the model, we introduce an auxiliary parameter as \textbf{electronic gain} \( e \), and select the degradation parameters with the following operation:
\begin{equation}
  \begin{aligned}
    &\log(e') \sim \mathcal{N}(0, e), \; \mathbf{K'} = e' \cdot \mathbf{K},&\\
    &\mathbf{D'} = e' \cdot \mathbf{D}, \; \bm{\sigma}_{\text{read}}' = e' \cdot \bm{\sigma}_{\text{read}},&
  \end{aligned}
  \label{eq:8}
\end{equation}
where \( \mathbf{K'} \), \( \mathbf{D'} \), and \( \bm{\sigma}_{\text{read}}' \) are the selected parameter values. By substituting HSI into the calibrated degradation model, realistic IHI data can be readily generated. Further details of the degradation model are in the Supplemental Document.

\subsection{IHRUT Network}
\label{sec:3.3}

\noindent \textbf{Degradation-guided Unfolding Architecture.}
Deep unfolding for HSI reconstruction integrates deep denoising networks with model-driven iterations. They model the imaging process with operations as \( \mathbf{y} = \mathbf{F}\mathbf{x} + \mathbf{n} \) (with \( \mathbf{F} \) as the imaging matrix and \( \mathbf{n} \) as noise) and formulate the following optimization problem:
\begin{equation}
  \begin{aligned}
    \hat{\mathbf{x}} = \min \frac{1}{2} \| \mathbf{y} - \mathbf{F}\mathbf{x} \|^2 + \tau J(\mathbf{x}).
  \end{aligned}
  \label{eq:9}
\end{equation}

Typical unfolding methods, such as Generalized Alternating Projection (GAP) and Proximal Gradient Descent (PGD)~\cite{GAPNet,RDLUF}, decompose the above optimization problem into data and prior subproblems, resulting in iterative updates as:
\begin{equation}
  \begin{aligned}
    &\mathbf{z}_{k+1} = \mathbf{x}_k + \bm{\rho} \mathbf{F}^T (\mathbf{y} - \mathbf{F}\mathbf{x}_k),&\\
    &\mathbf{x}_{k+1} = P(\mathbf{z}_{k+1}), &
  \end{aligned}
  \label{eq:10}
\end{equation}
where \( \bm{\rho} \) is a scale-invariant linear transformation that depends on the specific iterative method, and \( P \) denotes a denoising network.

For IHI, since the discrete FT can be represented as a matrix, the degradation model can also be transformed and introduced into Eq.~\ref{eq:10}. Specifically, by eliminating the signal-independent terms in the degradation model, let \( \mathbf{F}\mathbf{x} = \mathbf{K} \odot \mathbf{M} \odot \mathcal{F}\{\mathbf{A} \odot \mathbf{x} \} \). Given that \( \mathbf{F} \) is invertible, let \( \mathbf{F}' \) be the inverse of \( \mathbf{F} \), and define \( \bm{\alpha} = \bm{\rho} \mathbf{F}^T \mathbf{F} \). The iterative updates can then be rewritten as:
\begin{equation}
  \begin{aligned}
    &\mathbf{z}_{k+1} = \mathbf{x}_k + \bm{\alpha} \mathbf{F}' (\mathbf{y} - \mathbf{F}\mathbf{x}_k),&\\
    &\mathbf{x}_{k+1} = P(\mathbf{z}_{k+1}). &
  \end{aligned}
  \label{eq:11}
\end{equation}

The above formulation transforms the IHI problem and allows the unfolding methods for HSI reconstruction to be extended to IHI reconstruction. Our network is structured as shown in Fig.~\ref{fig:3a}. The network mainly consists of two submodules: Data and Prior, with Stripe-Pattern Enhancement (SPE) attention being the core mechanism. 

\noindent \textbf{Stripe Pattern Enhancement (SPE) Attention.}
In view of the stripe pattern characteristic, the degradation of IHI results exhibit strong correlations within and significant differences across \( (H \times 1 \times 1) \)-sized stripes. Conventional spectral mappings like convolution or channel attention cannot independently model and eliminate the spectral characteristics across stripes. SPE is designed to address this issue by independently modeling stripe-wise spectral characteristics in  the attention mechanism. As shown in Fig.~\ref{fig:3d}, SPE maps input \( \mathbf{x} \) via a depth-separable convolution embedding subnet \( C_1 \). Subsequently, \( H \)-direction average pooling and subnet \( C_2 \) are applied to leverage the low-rank characteristics of HSI and obtain a stripe-pattern attention map of size \( W \times d_{\text{in}} \). The feature from \( C_1 \) is further rescaled by the attention map and mapped to dimension \( d_{\text{out}} \) via a convolution, as:
\begin{equation}
  \begin{aligned}
    &\text{Attn} = C_2 \left( \text{AvgPool}_H \left( C_1(\mathbf{x}) \right) \right),&\\
    &\text{SPE}(\mathbf{x}) = \text{Conv}_{1 \times 1} \left( C_1(\mathbf{x}) \odot \left( \text{Attn} \right) \right).&
  \end{aligned}
  \label{eq:12}
\end{equation}

The design of SPE provides adaptability in handling stripe patterns and is applied across various parts of IHRUT.

\noindent \textbf{Adaptive Momentum Data Module.}
In the data module of IHRUT, we introduce two branches: projection and momentum. The projection branch uses the residual \( (\mathbf{y} - \mathbf{F}\mathbf{x}_k) \) for fidelity optimization, while the momentum branch, inspired by momentum gradient descent methods~\cite{FISTA, denoisingDU}, fuses iterative history across stages via \( (\mathbf{x}_k - \mathbf{x}_{k-1}) \) to enhance robustness. 
Subsequently, during the fusion of \( \mathbf{x}_k \) with both branches, we introduce a stripe-adaptive learning approach that involves two steps with SPE integration. As shown in Fig.~\ref{fig:3b}, input \( \mathbf{y} \) is mapped by a global \( \text{SPE}_0 \) (\( d_{\text{in}}=L \), \( d_{\text{out}}=K \)) for weighting to the projection at each stage. The fusion of branches \( \mathbf{r}_k \) is further weighted by internal \( \text{SPE}_k \) and combined with \( \mathbf{x}_k \) to produce \( \mathbf{z}_k \):
\begin{equation}
  \begin{aligned}
    &\mathbf{r}_k = \text{SPE}_0(\mathbf{y}) \cdot \mathbf{F}'(\mathbf{y} - \mathbf{F} \mathbf{x}_{k}) + (\mathbf{x}_{k} - \mathbf{x}_{k-1}),&\\
    &\mathbf{z}_k = \mathbf{x}_k + \text{SPE}_k (\mathbf{r}_k). &
  \end{aligned}
  \label{eq:13}
\end{equation}

This process assigns adaptive stripe-wise weights to both the projection and momentum branches, endowing the module with sufficient flexibility.

\noindent \textbf{Lightweight Cross-stage Prior Module.}
Originating from a U-shaped structure, our lightweight prior module replaces layer-wise dimension doubling with uniform dimensionality across layers, as shown in Fig.~\ref{fig:3c}. Additionally, to mitigate information loss during iterations, we introduced a cross-stage feature fusion mechanism. This mechanism integrates the features \( \textbf{f}_k \) with the downsampled features of the current stage, replacing the encoder in the U-shaped network. The decoders (IHRTrans) employ spatial-spectral transformer blocks with window-based multi-head self-attention (W-MSA) and SPE for adaptive stripe pattern correction and denoising. The output is then used as \( \textbf{f}_{k+1} \) and fed into the subsequent stage. The design of the prior module provides adequate performance while significantly reducing complexity (requiring only 50\% of the computational cost and parameters compared to recent U-shaped transformer-based methods~\cite{PADUT,RDLUF}, as shown in Tab.~\ref{tab:exp1}).

%% file: sec/_4_experiments.tex
\section{Experiments}
\label{sec:experiments}

\subsection{Datasets and Settings}
\label{sec:4.1}

In this section, we briefly introduce the experimental setup. All datasets (interferometric and spectral) are preprocessed to conform to the standard LASIS data format shown in Tab. \ref{tab:format}. The detailed implementation and data preparation process are described in the Supplemental Document.

\noindent \textbf{Synthetic Datasets.}
We prepare two HSI source datasets for simulation, namely HSOD-BIT {\cite{DMSSN}} and Houston. HSOD-BIT is a multi-spectral HSI dataset consists of 319 fine-quality HSIs originating from objective detection. The dataset possess suitable wavelength range (400-1000 nm) and a high resolution of 1240$\times$1680$\times$200. We randomly selected and preprocessed 50 images, generating 276 patches of 256$\times$256$\times$70 for trainnig and 6 scenes for testing. Houston is a commonly-used HSI in remote sensing of size 349$\times$1905 $\times$144 in 380-1050 nm, which is used as one scene for cross-domain test of methods. The size of 7 testing scenes is 256$\times$2048$\times$70. 

\noindent \textbf{Real LASIS Dataset.} 
We use absolute calibration data captured by the LASIS instrument (mentioned in Sec. \ref{sec:3.2}) as our testing dataset. The dataset includes real interferograms captured in uniform light and corresponding ideal HSIs as the ground truth. The formats of all real data are shown in Tab. \ref{tab:format}, with a size of \( H \)=120.
\begin{table}[h!]
    \centering
    \fontsize{7.5pt}{11pt}\selectfont
    \begin{tabular}{c|c|c|c}
        \toprule 
        \rowcolor{gray!20}
        & HSI & wavenumber & interferogram \\
        \hline
        width&\multicolumn{3}{c}{$W$=2048}\\
        \hline
        channels & $\Lambda=$70 & $N$=221 & $L$=256 \\
        \hline
        center & -&- &$l=$35\\
        \hline
        range & [450,900] nm & [0,0.0034] nm$ ^ {-1}$ & [-5140.8,32313.7] nm \\
        \hline
        resolution & 6.52nm & - & 146.88nm \\
        \hline
    \end{tabular}
    \caption{The standard data format of the LASIS instrument in our research. All data used in the IHI reconstruction experiments require preprocessing to align with this format.}
    \label{tab:format}
\end{table}
\begin{table*}[h!]
    \centering
    \fontsize{7.5pt}{9.5pt}\selectfont 
    \begin{tabular}{*{3}{c}|*{7}{c}|c|*{4}{c}}
        \toprule
        \rowcolor{gray!20}
        &&& \multicolumn{7}{c|}{\textbf{HSOD-BIT}} & \multicolumn{1}{c|}{\textbf{Houston}} & \multicolumn{4}{c}{\textbf{Calibration (Real)}} \\
        \cline{4-10}\cline{11-11}\cline{12-15}
        \rowcolor{gray!20}
        \multirow{-2}{*}{\textbf{Algorithms}}&\multirow{-2}{*}{\textbf{parameters}}&\multirow{-2}{*}{\textbf{GFLOPs}}& s1 & s2& s3 &s4 &s5 &s6 &avg &s1 &s1 & s2& s3& avg  \\
        \hline
        \multirow{2}{*}{traditional}& \multirow{2}{*}{-} & \multirow{2}{*}{-} & 21.38& 21.66 & 20.60 & 18.96& 23.27& 21.67& 21.26& 20.55&31.95&26.41&22.79&27.05\\
        &&& 0.906& 0.899 & 0.908 & 0.888& 0.922& 0.922& 0.908& 0.927&0.930&0.924&0.901&0.918 \\
        \hline
        \multirow{2}{*}{F'}& \multirow{2}{*}{-} & \multirow{2}{*}{-} & 25.73& 26.29 & 26.45 & 24.86& 27.61& 26.53& 26.25& 25.52&31.33&30.10&28.92&30.12 \\
        &&& 0.848& 0.832 & 0.858 & 0.818& 0.864& 0.889& 0.851& 0.915&0.847&0.807&0.765&0.806 \\
        \hline
        \multirow{2}{*}{DRUNet}& \multirow{2}{*}{32.73M} & \multirow{2}{*}{148.63} & 33.00& 35.57 & 35.13 & 37.17& 35.32& 34.58& 35.13& 30.99&37.30&32.05&28.56&32.64 \\
        &&& 0.977& 0.968 & 0.976 & 0.977& 0.977& 0.983& 0.976& 0.985&0.984&0.973&0.958&0.972 \\
        \hline
        \multirow{2}{*}{Restormer}& \multirow{2}{*}{26.18M} & \multirow{2}{*}{146.68} & 27.51& 29.40 & 29.00 & 28.15& 29.61& 27.23& 28.48& 25.54&33.73&28.71&24.95& 29.13 \\
        &&& 0.915& 0.899 & 0.919 & 0.888& 0.925& 0.933& 0.913& 0.944&0.908&0.857&0.804&0.856 \\
        \hline
        \multirow{2}{*}{SCUNet}& \multirow{2}{*}{9.74M} & \multirow{2}{*}{49.16} & 32.94& 35.81 & 35.03 & 37.21& 35.05& 35.05& 35.18& 31.57&37.13&31.50&28.42&32.35 \\
        &&& 0.974& 0.966 & 0.974 & 0.975& 0.975& 0.981& 0.974& 0.985&0.981&0.965&0.948&0.965 \\
        \hline
        \multirow{2}{*}{SST}& \multirow{2}{*}{4.16M} & \multirow{2}{*}{274.60} & 26.81& 30.11 & 29.88 & 28.17& 30.79& 27.10& 28.81& 25.09&33.17&27.77&25.41&28.78 \\
        &&& 0.937& 0.928 & 0.944 & 0.923& 0.948& 0.950& 0.938& 0.956&0.935&0.896&0.862&0.897 \\
        \hline
        \multirow{2}{*}{SERT}& \multirow{2}{*}{1.45M} & \multirow{2}{*}{92.64} & 33.53& 36.25 & 35.40 & 35.55& 35.95& 34.21& 35.15& 31.76&37.94&34.37&30.54&34.28 \\
        &&& 0.946& 0.939 & 0.954 & 0.939& 0.956& 0.962&0.949& 0.970& 0.949&0.919&0.885&0.918 \\
        \hline
        \multirow{2}{*}{GAP-net}& \multirow{2}{*}{4.39M} & \multirow{2}{*}{77.23} & 19.74& 19.58 & 19.17 & 21.23 & 20.26& 19.65& 19.94& 18.58&21.19&19.30&17.08&19.19 \\
        &&& 0.595& 0.563 & 0.635 & 0.563& 0.612& 0.647& 0.603& 0.656&0.610&0.467&0.344&0.474 \\
        \hline
        \multirow{2}{*}{PADUT-5}& \multirow{2}{*}{13.86M} & \multirow{2}{*}{201.59} & 40.75& 37.67 & 36.33 & 38.49& 37.65& 35.76& 38.34& 31.66&34.35&29.25&25.75&29.78 \\
        &&& 0.988& 0.976 & 0.982 & 0.986& 0.984& 0.989& 0.984& 0.991&0.988&0.985&0.980&0.984 \\
        \hline
        \multirow{2}{*}{RDLUF-3}& \multirow{2}{*}{9.29M} & \multirow{2}{*}{155.72} & 43.52& 41.47 & 41.00 & 44.13& 42.74& 43.34& 42.70& 38.46&40.95&37.54&34.34&37.61 \\
        &&& 0.983& 0.975 & 0.982 & 0.980& 0.984& 0.987& 0.982& 0.991&0.985&0.977&0.965&0.976 \\
        \hline
        \multirow{2}{*}{MAUN-7}& \multirow{2}{*}{4.78M} & \multirow{2}{*}{165.92} & 44.26& 41.63 & 41.46 & 45.34& 43.38& 43.93& 43.33& 39.72&40.96&39.18&34.30&38.15 \\
        &&& 0.990& \textbf{0.982} & \textbf{0.987} & \textbf{0.989}& \textbf{0.988}& 0.991& \textbf{0.988}& \textbf{0.995}&\textbf{0.995}&\textbf{0.994}&\textbf{0.988}&\textbf{0.992} \\
        \hline
        \rowcolor{gray!20}
        & & & 45.12& 42.82 & 42.45 & 46.42& 43.82& 44.85& 44.25& 39.73&\textbf{42.40}&40.51&35.47&39.46 \\
        \rowcolor{gray!20}
        \multirow{-2}{*}{\cellcolor{gray!20}IHRUT-5}&\multirow{-2}{*}{3.20M}&\multirow{-2}{*}{106.36}& \textbf{0.991}& \textbf{0.982} & \textbf{0.987} & 0.988& 0.987& 0.991& \textbf{0.988}& 0.994&\textbf{0.995}&0.993&\textbf{0.988}&\textbf{0.992} \\
        \hline
        \rowcolor{gray!20}
        & & & \textbf{45.25}& \textbf{42.64} & \textbf{42.46} & \textbf{46.56}& \textbf{43.98}& \textbf{44.91}& \textbf{44.30}& \textbf{39.86}&42.13&\textbf{40.66}&\textbf{36.04}&\textbf{39.61} \\
        \rowcolor{gray!20}
        \multirow{-2}{*}{\cellcolor{gray!20}IHRUT-7}&\multirow{-2}{*}{4.42M}&\multirow{-2}{*}{145.80}& \textbf{0.991}& \textbf{0.982} & \textbf{0.987} & \textbf{0.989}& \textbf{0.988}& \textbf{0.992}& \textbf{0.988}& \textbf{0.995}&\textbf{0.995}&\textbf{0.994}&\textbf{0.988}&\textbf{0.992} \\
        \hline
    \end{tabular}
    \caption{Comparison of reconstruction methods on (a) 6 synthetic scenes of HSOD-BIT. (b) 1 synthetic scene of Houton. (c) 3 real scenes of Calibration datasets with average values of 1000, 2000, and 3000. PSNR/SSIM (upper and lower entry in each cell, respectively) and complexity of different methods are shown. IHRUT demonstrates low complexity and superior performance.}
    \label{tab:exp1}
    \vspace{-10pt}
\end{table*}

\noindent \textbf{Implementation.} Our degradation model and the proposed IHRUT network, along with other comparative methods, are implemented based on Numpy and PyTorch. The learning-based methods are trained on a single RTX 3090 with the Adam optimizer, where \( \beta_1 \)=0.9 and \( \beta_2 \)=0.999. The total number of training epochs for all methods is 300, using a cosine annealing scheduler and linear warm-up. The batch size is 1, and the initial learning rate is 1e-4. The optimization objective of learning is the L1 loss.

\noindent \textbf{Evaluation Metrics.} The reconstruction quality is evaluated by peak signal-to-noise ratio (PSNR) and structural similarity index (SSIM), while the complexity of the model is assessed by GFLOPs (evaluated on training patches) and the number of parameters (in millions).
\begin{table}[h!]
    \centering
    \fontsize{7.5pt}{9pt}\selectfont 
    \begin{tabular}{c|c|c|c}
        \toprule
        \rowcolor{gray!20}
        &\multicolumn{2}{c|}{Parameters} & Metrics \\
        \cline{2-4}
        \rowcolor{gray!20}
        \multirow{-2}{*}{Model}& Optical& Electronic & PSNR / SSIM\\
        \hline
        $m_1$&Trad.&-&27.05 / 0.918\\
        \hline
        $m_2$&$\mathbf{A},\bm{\beta}$&-&26.20 / 0.863\\
        $m_3$&$\mathbf{A},\bm{\beta},\bm{M}$&-&33.96 / 0.839\\
        \hline
        $m_4$&$\mathbf{A},\bm{\beta},\bm{M}$&Gaussian Noise&37.95 / 0.965\\
        
        $m_5$&$\mathbf{A},\bm{\beta},\bm{M}$&Standard Noise&38.82 / 0.987\\
        \hline
        $m_6$&$\mathbf{A},\bm{\beta},\bm{M}$&Our Noise (pixel-wise)&39.28 / 0.988\\
        \hline 
        ours&$\mathbf{A},\bm{\beta},\bm{M}$&Our Noise + $e$&\textbf{39.46} / \textbf{0.992}\\ 
        \hline
    \end{tabular}
    \caption{Validation of degradation model on real dataset. Among the models, $m_4$ is equipped with Gaussian white noise \cite{Gaussian1} for, and $m_5$ with standard noise model ($\mathbf{N}_{shot},\mathbf{N}_{read}$) for comparison. Our model achieve the optimal performance.}
    \label{tab:exp2}
    \vspace{-10pt}
\end{table}

\subsection{Degradation Model Validation}
To verify the effectiveness of our degradation model, we train the 5-stage IHRUT network with data simulated from our model along with 6 comparative models, and analyze the reconstruction performance on real testing dataset. Among these models, model \( m_1 \) is used for guiding traditional reconstruction. The models \( m_2 \) and \( m_3 \) serves as ablations of our optical parts (such as $\mathbf{M}$). \( m_4 \) and \( m_5 \) are equipped with Gaussian white noise and standard noise model ($\mathbf{N}_{shot},\mathbf{N}_{read}$) for comparison. With our design of pixel-wise noise modeling for the stripe-pattern noise distribution, \( m_6 \) outperforms standard noise model by 0.64dB. Furthermore, the introduction of \( e \) effectively enhances robustness (+0.18 dB). The comparative results validates the rationality and accuracy of our degradation model structure.

\subsection{Reconstruction Experiments}

\noindent \textbf{Comparative Methods.}
In the experiments, the proposed IHRUT is compared with three categories of methods: (1) traditional methods such as the LASIS restoration method from the Xi'an Institute of Optics and Mechanics, and \( \mathbf{F} \)' mentioned in Sec. \ref{sec:3.3}. (2) End-to-End (E2E) networks without priors in learning, which integrate FT as a preprocessing step with general restoration networks like DRUNet, Restormer, and SCUNet \cite{drunet,Restormer,scunet}, as well as HSI denoising networks like SST and SERT \cite{SST,SERT}. (3) Deep Unfolding HSI reconstruction methods adapted for IHI such as GAP-Net, PADUT, RDLUF, and MAUN \cite{GAPNet,PADUT,RDLUF,MAUN}. For each unfolding method, the results shown in Tab. \ref{tab:exp1} are for the optimal number of stages, while the analysis of different stage numbers can be found in the Supplemental Document.
\begin{figure*}
    \centering
    \includegraphics[width=\textwidth]{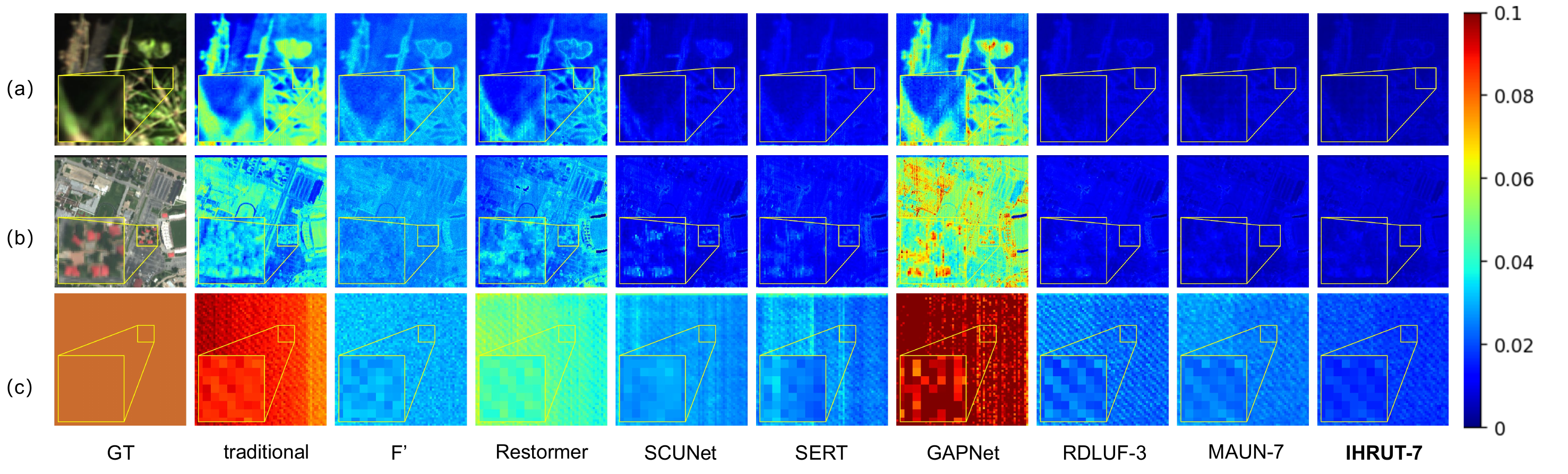}
        \caption{Comparison of reconstructed HSI with error maps (10$\times$ amplified). We select Synthetic (a) Scene 2 from HSOD-BIT and (b) Scene 1 from Houston cropped by 256$\times$256, along with Real (c) Scene 3 from Calibration dataset cropped by 64$\times$64. The error map exhibits the spectral absolute error of IHI reconstruction results by 8 algorithms and IHRUT with 7 stages. The region within the box is chosen for analysis of the reconstructed spectra and zoom in for a more detailed examination. }
        \label{fig:exp1}
        \phantomsubcaption\label{fig:exp1a}
        \phantomsubcaption\label{fig:exp1b}
        \phantomsubcaption\label{fig:exp1c}
    \end{figure*}

\noindent \textbf{Simulated IHI Reconstruction.}
Tab. \ref{tab:exp1} presents the quantitative results of all methods on 6 scenes of HSOD-BIT and 1 scene of Houston. Our IHRUT achieves the best performance across all metrics, showing an average advantage of 0.85 dB over MAUN. The Houston remote sensing scene poses a challenge to the cross-domain generalization ability of reconstruction algorithms. Some methods, such as PADUT, show degraded performance, while IHRUT maintains superior performance with 0.14 dB than MAUN. As shown in Fig. \ref{fig:exp1a} and Fig. \ref{fig:exp1b}, the visualization of the reconstruction errors further validates the quantitative results. E2E methods tend to leave noticeable color differences and stripe noises, whereas IHRUT achieves the smallest reconstruction errors, confirming its effectiveness in detail enhancement.

\noindent \textbf{Real IHI Reconstruction.} 
Tab. \ref{tab:exp1} displays the quantitative results on 3 scenes of the real calibration dataset. For E2E methods that do not incorporate degradation priors, even the better-performing SCUNet and SERT exhibit a gap of approximately 4 dB compared to recent unfolding methods, demonstrating the challenges in addressing the degradation process. Among the unfolding methods, our IHRUT shows a significant advantage, with an average PSNR improvement of approximately 1.4 dB compared to MAUN. In the visualization results shown in Fig. \ref{fig:exp1c}, most reconstruction methods retain varying degrees and types of noise, while IHRUT outperforms other methods in terms of noise removal.

\noindent \textbf{Complexity.} As shown in Tab. \ref{tab:exp1}, IHRUT benefits from its lightweight design and exhibits lower complexity compared to other unfolding methods. Particularly, the 5-stage version of IHRUT achieves a low complexity while maintaining high performance.

\vspace{-5pt}
\begin{table}[h!]
    \centering
    \fontsize{7.5pt}{9pt}\selectfont 
    \begin{tabular}{c|*{3}{c}|*{2}{c}|c}
        \toprule 
        \rowcolor{gray!20}
        &\multicolumn{3}{c|}{Data} & \multicolumn{2}{c|}{Prior} & \multicolumn{1}{c}{Metrics} \\
        \cline{2-4}\cline{5-6}\cline{7-7}
        \rowcolor{gray!20}
        \multirow{-2}{*}{baseline}& $\text{SPE}_0$ & $\text{SPE}_k$ & Mom. & C-Stg. & SPE & PSNR/SSIM\\
        \hline
        \checkmark&&&&&&34.37 / 0.934\\
        \hline
        \checkmark&\checkmark&&&&&38.73 / 0.982\\
        \checkmark&\checkmark&\checkmark&&&&39.10 / 0.983\\
        \hline
        \checkmark&\checkmark&\checkmark&\checkmark&&&39.12 / 0.984\\
        \checkmark&\checkmark&\checkmark&\checkmark&\checkmark&&39.35 / 0.985\\
        \checkmark&\checkmark&\checkmark&\checkmark&\checkmark&\checkmark&\textbf{39.46} / \textbf{0.992} \\
        \hline
    \end{tabular}
    \vspace{-8pt}
    \caption{Break down ablation of IHRUT (5-stages) on Calibration dataset. CA denotes replacing SPE with channel attention.}
    \label{tab:exp3}
\end{table}
\vspace{-10pt}

\subsection{Ablation Study}
Under the validated overall efficacy of the proposed framework, ablation experiments are conducted in two stages. First, a comprehensive breakdown ablation is performed on every major component integrated in IHRUT. As reported in Table \ref{tab:exp3}, the SPE mechanism emerges as the most critical contributor, delivering a performance gain exceeding 5 dB, whereas the momentum mechanism and cross-stage fusion play auxiliary roles in boosting quality of reconstruction.

\begin{figure}
  \centering
  \vspace{-8pt}
  \includegraphics[width=\linewidth]{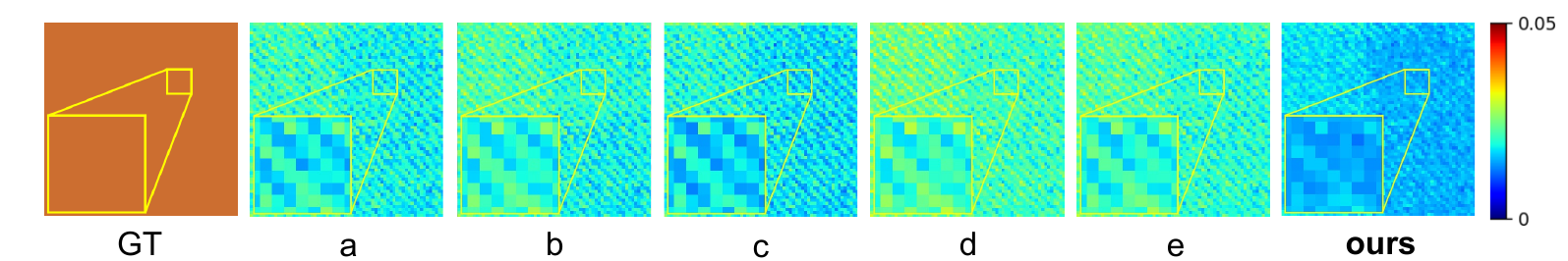}
  \setlength{\abovecaptionskip}{-10pt}
  \caption{Error map (20$\times$) visualization of Tab. \ref{tab:exp4}.}
  \label{fig:exp2}
  \vspace{-12pt}
\end{figure}

Subsequently, we undertake a detailed comparative analysis of our SPE and Momentum Branch against channel-attention (Tab. \ref{tab:exp4} a,b,c) and existing momentum branches (Tab. \ref{tab:exp4} d,e). As illustrated in Tab. \ref{tab:exp4} and Fig. \ref{fig:exp2}, the stripe-wise attention introduced by SPE yields superior flexibility, and the two successive weighting operations within our momentum pathway further elevate the overall performance.


\begin{table}[h!]
  \centering
  \fontsize{6.5pt}{8pt}\selectfont 
  \begin{tabular}{c|ccc}
      \hline

      \hline
       Data Module&a&b&c\\
      \hline
       $z_{k}-x_k$&$\text{CA}(\text{CA} \cdot \delta$+$m)$&$\text{CA}(\text{SPE}_{0}\cdot \delta$+$m)$&$\text{SPE}_{k}(\text{CA} \cdot \delta$+$m)$\\
       PSNR/SSIM&38.55/0.984&38.66/0.984&38.60/0.985\\
      \hline

      \hline
       Data Module&d (like \cite{FISTA,denoisingDU})&e&\textbf{ours}\\
      \hline
      $z_{k}-x_k$&$\text{CA}\cdot \delta$+$\text{CA}(m)$&$\text{SPE}_{k}\cdot \delta$+$\text{SPE}_{0}(m)$&$\text{SPE}_k(\text{SPE}_0 \cdot \delta$+$m)$\\
       PSNR/SSIM&38.31/0.983&38.40/0.985&\textbf{39.46/0.992}\\
      \hline

  \end{tabular}
  \vspace{-8pt}
  \caption{Analysis of the Data Modules with different components and momentum branches (e.g., $z_{k}-x_{k}$ in the table) in 5-stg IHRUT. Here, $\delta=\mathbf{y}-\mathbf{F} \mathbf{x}_k$ and $m=\mathbf{x}_{k}-\mathbf{x}_{k-1}$.}
  \label{tab:exp4}
\end{table}

\vspace{-10pt}

%% file: sec/_5_conclusion.tex
\section{Conclusion}
\label{sec:conclusion}

In this paper, we focus on enhancing the performance of IHI reconstruction through learning approach. We establish a simplified yet accurate degradation model and calibration method to extract degradation priors from IHI imaging. leveraging the model, we conduct synthesis for realistic data to tackle the lack of dataset. Based on degradation model, we design the IHRUT network. Within a low-complexity spatial-spectral unfolding architecture, it is optimized for the stripe characteristics of IHI degradation. In experiments involving training with simulated data and joint testing with simulated and real data, IHRUT demonstrates outstanding performance. We hope our work could provide inspiration for further researches on optical spectroscopy instruments.